\documentclass[12pt]{elsarticle}
\usepackage[numbers]{natbib} 

\usepackage{amsmath}
\usepackage{comment}
\usepackage{graphicx}
\usepackage{subfig}
\usepackage{caption}
\usepackage{gensymb}
\usepackage{setspace}
\usepackage{hyperref}
\usepackage{lineno}
\usepackage{amssymb}
\usepackage{algorithm}
\usepackage{algpseudocode}
\usepackage{xcolor}
\usepackage[T1]{fontenc}
\usepackage{svg}
\usepackage{float}

\newcommand*{\figref}[2][]{%
  \hyperref[{fig:#2}]{%
    Figure~\ref*{fig:#2}%
    \ifx\\#1\\%
    \else
      .#1%
    \fi
  }%
}

\journal{}

\title{High-End Space Electronics: \\ Active Shielding to Mitigate Catastrophic Single-Event Effects}

\author[1,2]{Yoav Simhony}

\author[3]{Alexander Segal}

\author[2]{Ofer Amrani}

\author[1]{Erez Etzion}

\affiliation[1]{organization={Tel Aviv University},addressline={Raymond and Beverly Sackler School of Physics and Astronomy}, city={Tel Aviv},postcode={69978}, country={Israel}}
    
\affiliation[2]{organization={Tel Aviv University},addressline={School of Electrical Engineering}, city={Tel aviv},postcode={69978}, country={Israel}}

\affiliation[3]{organization={Afeka College of Engineering},addressline={Unit of Mathematics}, city={Tel Aviv},postcode={6910721}, country={Israel}}

\begin{document}

\begin{frontmatter}

\begin{abstract}
Operating electronic systems in space environments presents significant challenges due to continuous exposure to cosmic, solar, and trapped radiation, which can induce catastrophic single-event effects. This paper introduces a novel nonintrusive mitigation apparatus designed to protect high-end commercial off-the-shelf electronics in space. The apparatus incorporates an array of real-time particle detectors coupled with a mitigation algorithm. Upon identifying potentially harmful particles, the system power cycles affected electronics, preempting permanent damage. The apparatus was evaluated using GEANT4 simulations, which were compared with empirical data from the "COTS-Capsule" experiment aboard the International Space Station, demonstrating strong agreement. Key results indicate that the system achieves a 95\% detection accuracy with a power cycle rate of once every seven hours per square centimeter of sensitive electronics. The COTS-Capsule represents a cost-effective, flexible solution for integrating modern, high-end, non-space-qualified electronics into a variety of space missions, addressing critical challenges in the new-space era.

\end{abstract}

\begin{keyword}
Electronic space systems, Single-event effects (SEE), Radiation mitigation, Particle detectors, International Space Station (ISS)

\PACS 0000 \sep 1111

\MSC 0000 \sep 1111
\end{keyword}

%%======================================%%
%%  Introduction                        %%
%%======================================%%
\end{frontmatter}

\section{Introduction}\label{sec1}

Space exploration and utilization have witnessed a paradigm shift in the last decade, coined "The new-space revolution" \cite{DENIS2020431, pelton2019space}. New commercial entities have provided advancements in launch vehicles and satellite technology, thus facilitating a broad range of space-oriented solutions. These include scientific areas such as planetary exploration, Meteorology, Astrophysics, and Biology, as well as engineering applications such as telecommunications and remote sensing, to name but a few \cite{MILLAN20191466}. However, the harsh space environment, characterized by radiation, vacuum, and micro-gravity, continues to pose numerous challenges for the design and operation of spacecrafts~\cite{8758870, maurer2008harsh}.

Understanding and addressing the unique environmental challenges of space is essential for the success of space missions and for the safety of those involved. Space environments significantly impact the performance and reliability of spacecraft systems. Scientists and engineers have been studying these key concerns since the dawn of the space age ~\cite{doi:10.1126/science.131.3398.385}, even before the launch of the first man-made satellite ~\cite{singer1956effect}. 

Since maintenance in space is intricate and, in many cases, impossible, designing electronic systems for space use is challenging as they must withstand  
the effects of space particle radiation, including cosmic, solar, and trapped particle radiation. Such particles can have two potentially detrimental impacts on non-space-grade electronics ~\cite{electronics10091008}. These include cumulative radiation effects such as total ionizing dose (TID) and single-event effects (SEEs) that result from the impact of a single particle ~\cite{velazco2019radiation}.

SEEs can be categorized into single-event upsets (SEUs) and catastrophic single-event effects (CSEEs). SEUs are temporary errors causing memory "bit-flips", on the other hand, CSEEs cause irreversible damage to sensitive electronic components by inducing relatively high current draw within the components \cite{bruguier1996single, johnston1996influence, 1208579, duzellier2005radiation}.

Spacecraft engineers have developed various strategies to address the challenges posed by particle radiation in space. One approach involves using space-grade electronics designed and qualified to tolerate %withstand 
the effects of radiation. These components may be fabricated using specialized technologies such as silicon on insulator (SOI)~\cite{schwank2003radiation}, which are typically immune to SEEs or may be designed with specific very large-scale integration (VLSI) libraries that enhance their radiation resistance~\cite{4674813, 6131334, maurer2008harsh}.

Another strategy involves commercial electronic components that have undergone space qualification testing and demonstrated good radiation tolerance. These components may be used in space systems in some cases, although they might require additional mitigation measures~\cite{velazco2019radiation, electronics10091008, GULDAGER2005279, 8093122, 819113}. 

A straightforward mitigation measure is passive shielding, which entails enclosing sensitive electronics in a protective enclosure. Such enclosures absorb low-energy particles, reducing the radiation flux that reaches the sensitive electronics and mitigating cumulative radiation effects. The drawback of passive shielding is that it increases the spacecraft's mass and launch cost, although advancements in launching services and reduced launch costs have diminished this concern. Nevertheless, passive shielding is generally ineffective in mitigating SEEs~\cite{8003007, maurer2008harsh, sinclair2013radiation}.

In general, SEUs are addressed utilizing error detection and correction techniques such as parity checking and error correction codes. While these solutions require processing and consume additional electric power, they are highly effective in mitigating SEUs~\cite{maurer2008harsh, nidhin2018review, tang2014soft, siegle2015mitigation, sinclair2013radiation, li2007software}.

When space-qualified electronics are susceptible to CSEEs, it is imperative to employ a mitigation strategy. CSEEs may occur when a single high atomic number and energy (HZE) ion (atomic number $Z>2$) strikes a sensitive electronic component. Depending on the particle's initial energy and atomic number, it may trigger a CSEE, resulting in an internal short circuit within the component. This, in turn, generates localized thermal heating and potentially causes permanent thermally-induced damage at the semiconductor component level~\cite{maurer2008harsh, velazco2019radiation, sinclair2013radiation}.

Typically, the duration of a CSEE progression before reaching %until 
thermal catastrophic failure is more than a millisecond. By promptly detecting a CSEE and interrupting its progress within a shorter time frame, it is feasible to prevent the event from escalating into a catastrophic failure~\cite{velazco2019radiation, maurer2008harsh, sinclair2013radiation}.

One way to address this challenge is by implementing a current monitoring circuit that detects an increase in the current draw of the sensitive component beyond a predefined threshold. When such an increase is detected, it initiates an immediate power cycle, which involves turning the electronic device off by disconnecting it from its power supply and then turning it back on. Current turn-off mitigates the potential damage induced by the otherwise CSEE. Unfortunately, this solution calls for the development of a dedicated electronic protection circuit for each sensitive component. It also requires radiation-testing each component and characterizing its sensitivity to particle radiation. Custom design of spacecraft electronics, restricted to utilizing only radiation-qualified electronics, is both time-consuming and expensive. Moreover, this method cannot be implemented when protecting high-current consumption or variable current consumption components, let alone complete electronic boards such as high-end computer processing boards~\cite{maurer2008harsh, velazco2019radiation, sinclair2013radiation}.

%%===============================================%%
%%  The COTS-Capsule CSEE Mitigation Scheme      %%
%%===============================================%%
%
\section{The COTS-Capsule CSEE Mitigation Scheme}

This paper proposes a novel approach to mitigating CSEEs in terrestrial Commercial Off-The-Shelf (COTS) electronics when used in space systems. We introduce the “COTS-Capsule”, a non-intrusive CSEE mitigation apparatus designed to protect terrestrial electronics in space.

This method offers several advantages over traditional approaches. Utilizing readily available, mass-produced electronics not only enhances the overall performance of space systems but also simplifies the integration of new space-oriented applications. With COTS-Capsule, existing terrestrial electronic boards can be deployed in space, minimizing the need for extensive space qualification and radiation-hardening processes. This approach significantly reduces the time and costs associated with space-specific board and system design. Furthermore, by facilitating easy integration and “last-minute” swapping of modern electronics, this approach supports more efficient and cost-effective design processes for space systems.

The COTS-Capsule utilizes an array of particle detectors that encapsulate CSEE-sensitive electronics, allowing real-time detection and characterization of particles traversing the electronics. A CSEE mitigation algorithm assesses the risk of each particle event and, if a potential CSEE is identified, initiates a power cycle of the sensitive electronics, preemptively neutralizing the threat and preventing permanent damage.

However, indiscriminate power cycling in response to all particle interactions would lead to excessive cycling rates, making it inefficient for practical use due to the vast number of benign particles in space. Therefore, the COTS-Capsule apparatus has been optimized to identify, characterize, and mitigate only potentially CSEE-inducing particles, ensuring that sensitive electronics operate reliably and with minimal interruptions in orbit.

\begin{figure}[H]%
\centering
\includegraphics[width=0.9\textwidth]{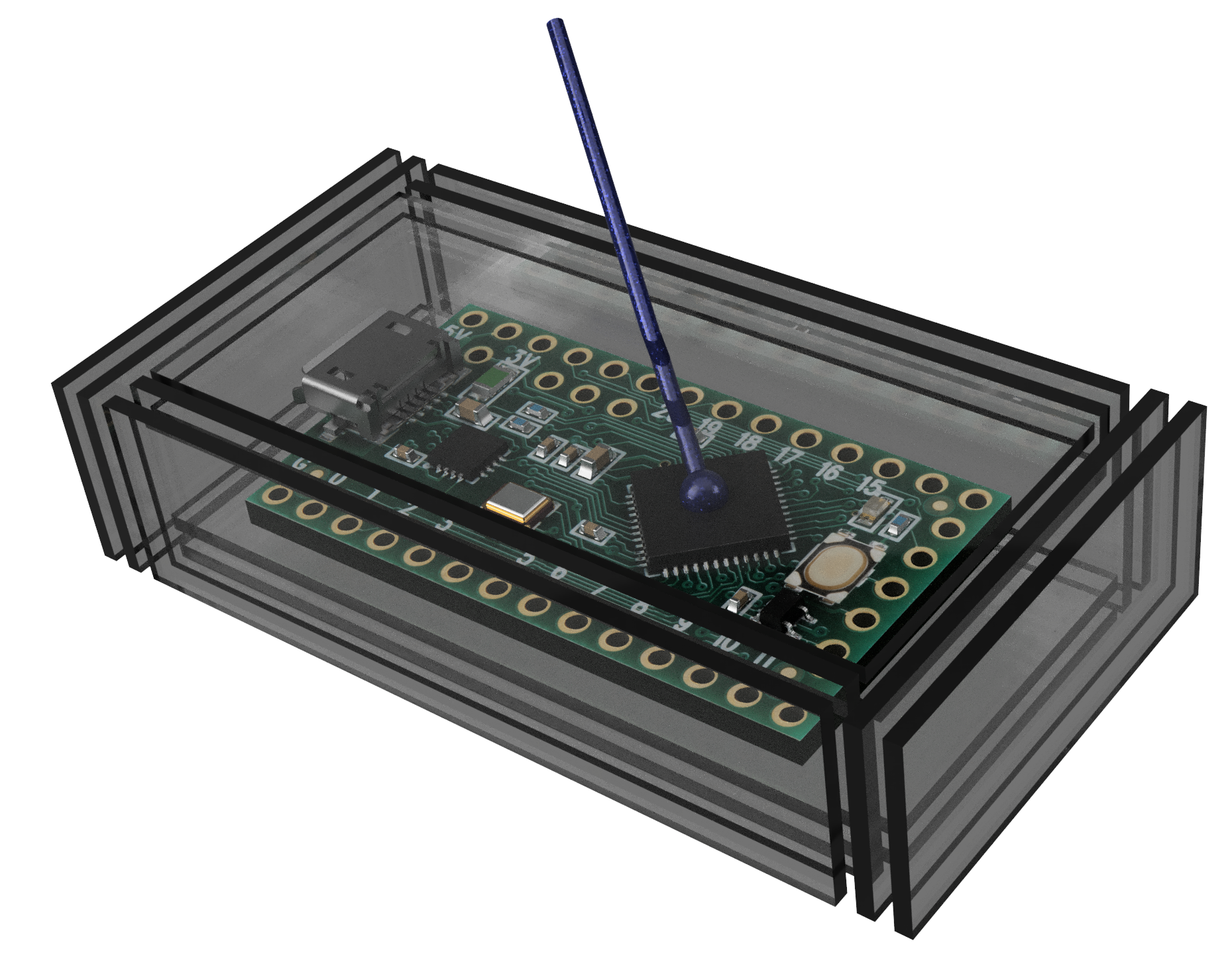}
\caption{Illustration of a COTS-Capsule apparatus. A CSEE-sensitive COTS electronic board (3D model of Teensy\textsuperscript \textregistered \ LC with permission) is encapsulated within a rectangular prism-shaped casing encompassing two particle detectors (clear) on each side of the prism's faces. A particle (blue ball) traverses the top two detectors and impinges a sensitive component. The particle track is also illustrated (blue track).}\label{fig:cots-capsule_concept_design}
\end{figure}

In practical implementation, the sensitive electronic board is enclosed within a rectangular prism-shaped casing with a particle-detector array on each of its six faces. Each face consists of %contains
two layers of position-sensitive detectors, totaling twelve detectors. With this configuration, any potentially CSEE-inducing particles traversing the sensitive electronics are detected by at least two of the COTS-Capsule’s position-sensitive detectors, as illustrated in \figref{cots-capsule_concept_design}.

Because CSEE-inducing particles, primarily ions with high atomic number and energy (HZE), %ions, 
generally follow near-straight trajectories, the apparatus can accurately determine a particle’s path by detecting its intersections with the position-sensitive detectors. This enables precise determination of whether the particle traversed any sensitive components, necessitating a power-cycle. Correspondingly, reducing the power cycle rate significantly, as CSEE-sensitive components typically occupy but a fraction of the total area of a printed circuit board (PCB).

Furthermore, by analyzing the energy deposition along a particle’s trajectory, the system can often differentiate between particle types or, at the very least, assess the potential for a CSEE. Notably, only a small subset of cosmic particles—specifically HZE ions—can induce CSEEs \cite{maurer2008harsh, 4333163, RAUCH20141444, sinclair2013radiation}. By isolating HZE ions and initiating power cycling exclusively for particles capable of triggering CSEEs, the COTS-Capsule significantly reduces the frequency of power cycles.

In this study, we developed and tested a CSEE mitigation algorithm based on three sequential steps:
\begin{enumerate}
    \item For each detected particle event, estimate the particle’s trajectory to assess if it intersected with any sensitive component;
    \item if a sensitive component was intersected, estimate the energy deposited by the particle at each detection layer;
    \item when the deposited energy indicates a potential CSEE risk, initiate a rapid power cycle to reset any triggered CSEE before permanent damage occurs.
\end{enumerate}

%%=============================================================%%
\section{The COTS-Capsule Emulator}
%%=============================================================%%

The COTS-Capsule apparatus was designed and validated using SRIM~\cite{ZIEGLER20101818} simulations, which facilitated the development and evaluation of the CSEE mitigation algorithm. The apparatus was subsequently emulated and validated in space environments through both empirical and simulation-based approaches.

\begin{itemize}
\item Empirical COTS-Capsule Emulator: The COTS-Capsule apparatus was emulated using measurements obtained from our dedicated spaceborne experiment~\cite{COTS-CAPSULE-SYSTEM} conducted onboard the International Space Station (ISS).
\item Simulated COTS-Capsule Emulator: The empirical performance was further validated by simulating the COTS-Capsule apparatus and mitigation algorithm within a modeled space environment. This simulation utilized CREME96~\cite{tylka1997creme96} physical radiation models and was implemented using the GEANT4 simulation toolkit~\cite{AGOSTINELLI2003250}.
\end{itemize}

The agreement between empirical measurements and simulation results confirms the COTS-Capsule’s viability and effectiveness as a CSEE mitigation scheme. The apparatus effectively protects sensitive electronics while maintaining a power-cycle rate optimized for a broad range of space missions.

%%=============================================================%%
%% Empirical COTS-Capsule Emulator %%
%%=============================================================%%

\subsection{Empirical On Orbit COTS-Capsule Emulator}

%%=============================================================%%
%% COTS-Capsule Spaceborne Mission
%%=============================================================%%

\subsubsection{COTS-Capsule Spaceborne Mission}

\begin{figure}[h!] %[h]%
\centering
\includegraphics[width=0.9\textwidth]{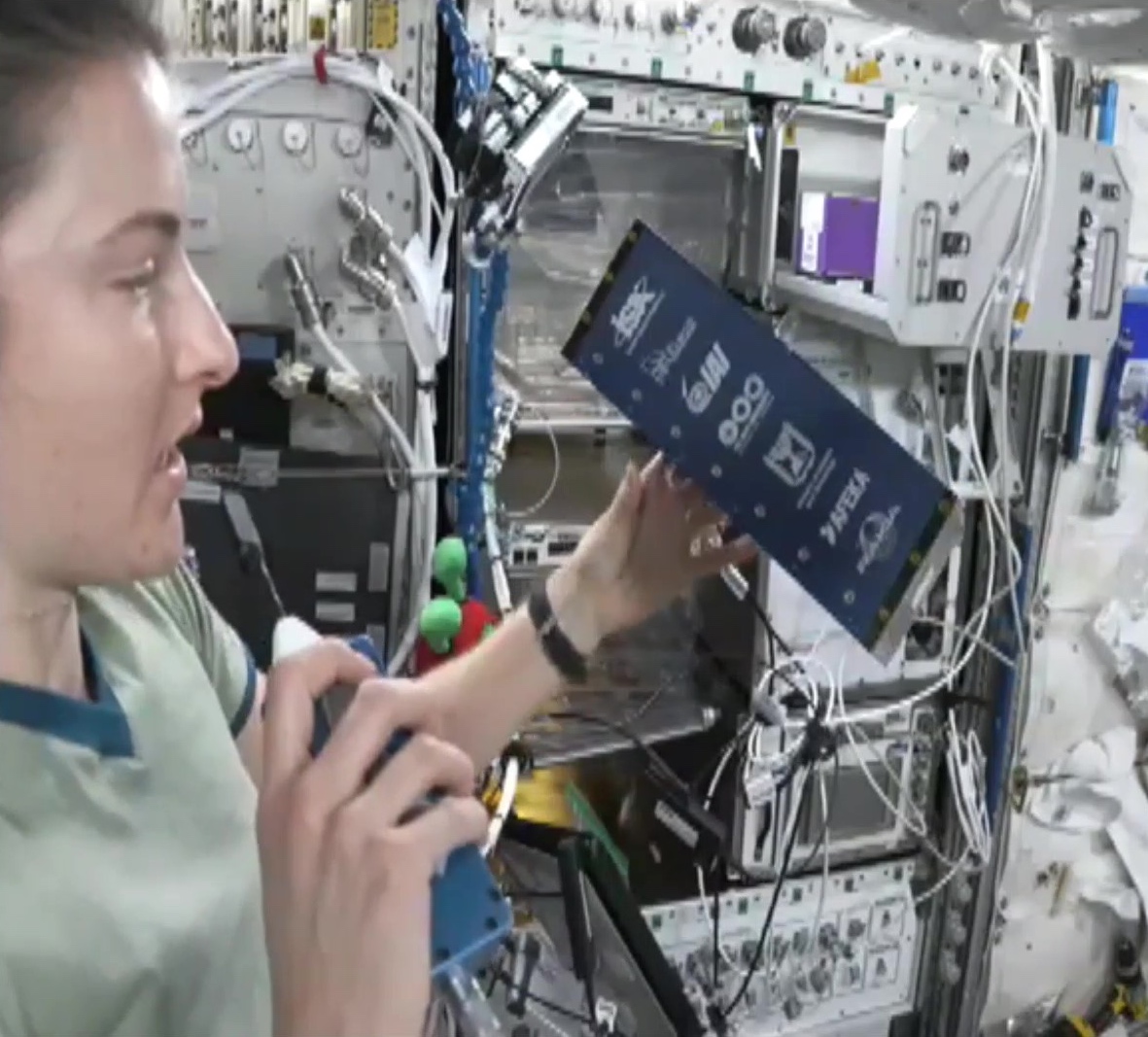}
\caption{Installation of the COTS-Capsule aboard the ISS, undertaken by NASA astronaut Kayla Barron (image courtesy of NASA/Voyager Space).}\label{fig:overlay}
\end{figure}

The COTS-Capsule experiment~\cite{COTS-CAPSULE-SYSTEM} was successfully deployed aboard the ISS, as shown in \figref{overlay}, with core measurements conducted in 2022. This experiment was instrumental in acquiring the data necessary to refine the design of the COTS-Capsule apparatus and to validate the operation of its core components in a space environment.

\begin{figure}[!h] %[h]%
\centering
\includegraphics[width=0.9\textwidth]{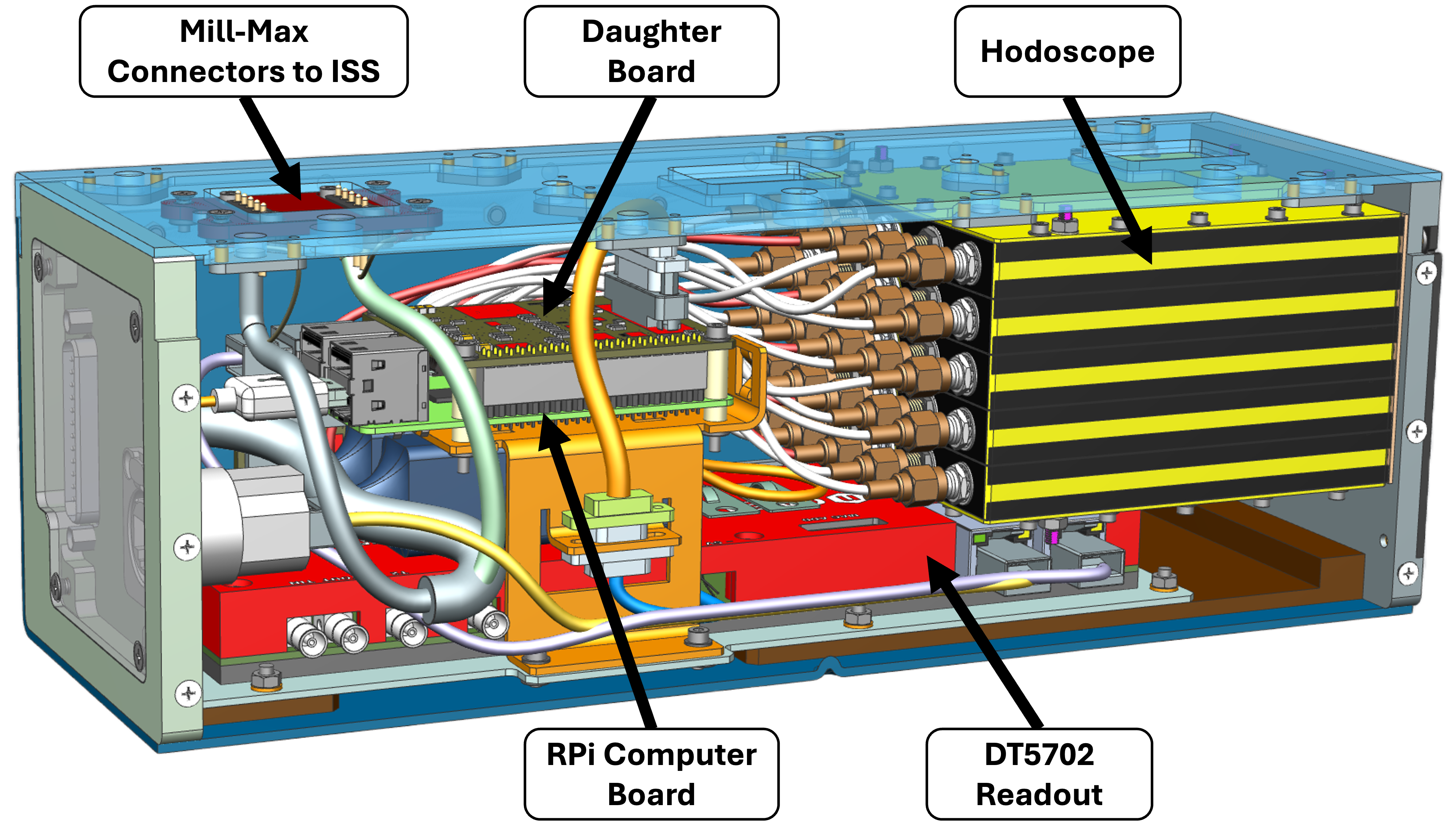}
\caption{CAD model of the COTS-Capsule spaceborne experiment that was launched and operated aboard the ISS. The ISS provides power and communication through two Mill-Max connectors at the top. On the right side, a five-layer hodoscope contains five vertically-stacked SSPDs~\cite{SSPD}, connected to a DAQ DT5702 via coaxial cables. The CAEN DT5702 readout electronics board~\cite{dt5702} is positioned at the bottom in red. The central section houses the onboard computer, which includes a Raspberry Pi (RPi) board~\cite{RaspberryPi} with an attached custom-designed daughter board. }\label{fig:complete_assembly}
\end{figure}

\figref{complete_assembly} shows the design of the spaceborne COTS-Capsule experiment aboard the International Space Station (ISS). On the right side of the payload, a five-layer hodoscope is designed to detect, track, and evaluate particle energy loss as they traverse the apparatus. The hodoscope incorporates five Scintillator-SiPM Particle Detectors (SSPD) ~\cite{SSPD}, developed specifically for the COTS-Capsule apparatus. At the bottom, highlighted in red, is the CAEN DT5702~\cite{dt5702} readout electronics board, while the center of the payload houses a Raspberry Pi (RPi)~\cite{RaspberryPi} computer with a custom daughter board.

During the course of the experiment, we successfully collected data from over half a billion particle events, encompassing particles from various sources such as solar radiation, cosmic radiation, and particles trapped within the Van Allen radiation belts~\cite{vanallen1959radiation}.
In addition, the system measured secondary particles generated by interactions between primary particles and ISS equipment and shielding. 

\begin{figure}[!htbp]
  \centering
  \subfloat[]{\includegraphics[width=1\textwidth]{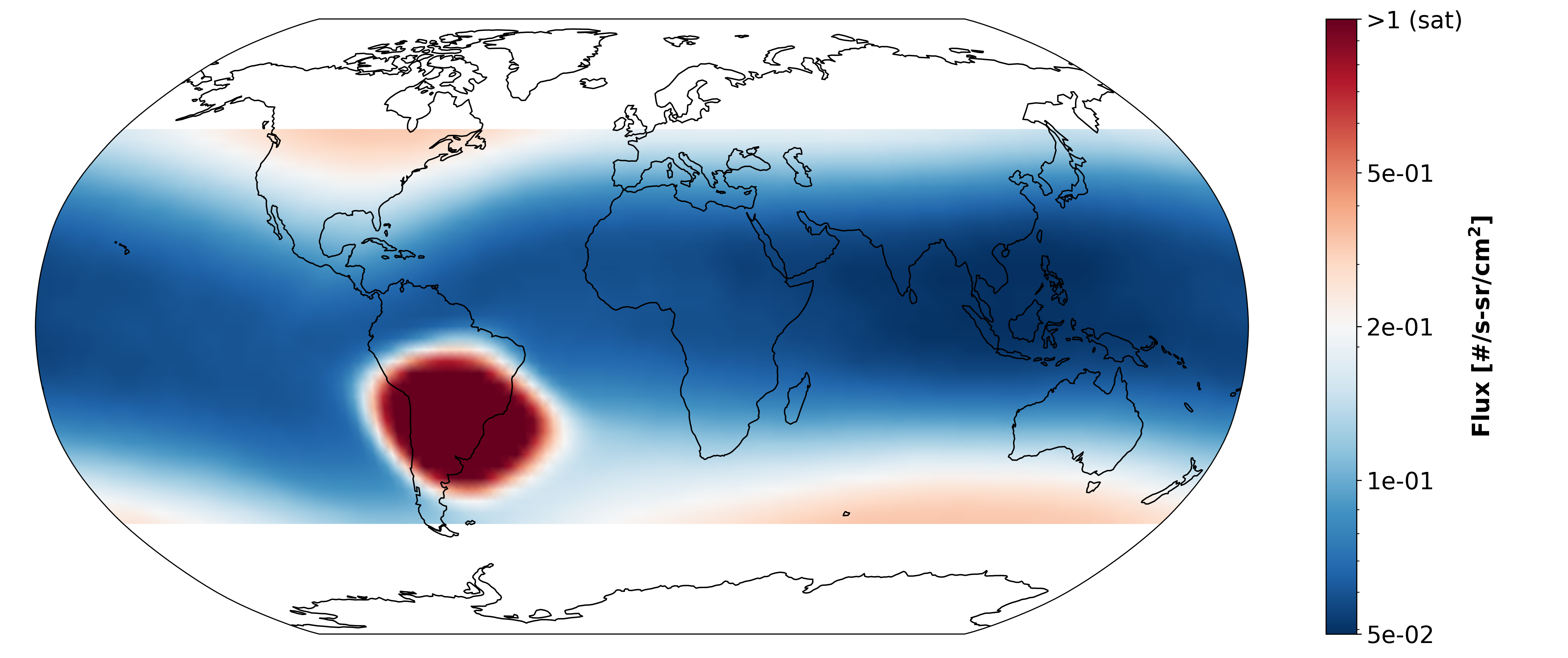}}
  \hfill
  \subfloat[]{\includegraphics[width=1\textwidth]{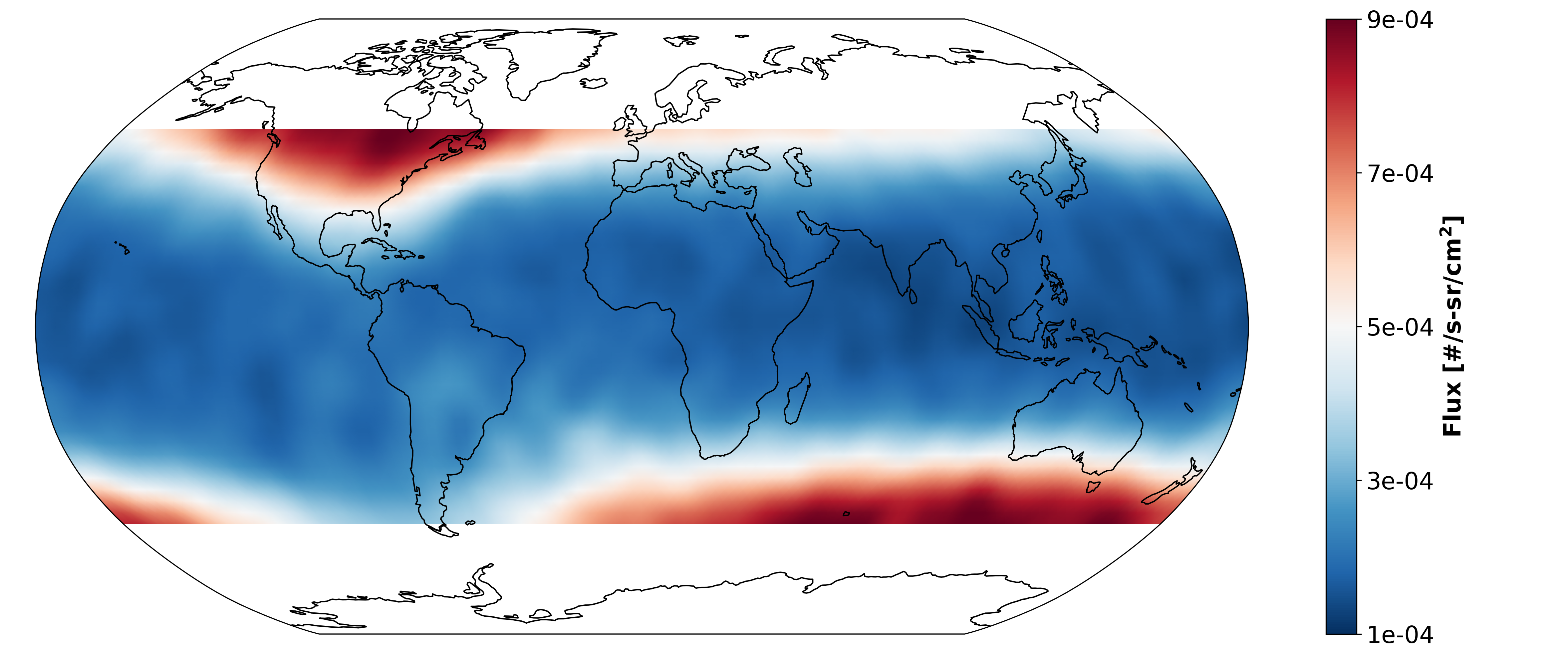}}
 \caption{
 The first results of the spaceborne COTS-Capsule experiment. Heat maps of the particle flux measured onboard the ISS (2022). (a) All particles detected by the COTS-Capsule. To improve the discernibility of flux variability across non-SAA locations, the particle flux depicted is truncated at a fixed threshold. (b) The HZE ion flux detected by the COTS-Capsule.
 }
  \label{fig:overall_vs_ions}
\end{figure}

The data collected, depicting the particle flux measured by our apparatus aboard the ISS, is presented as a heat map overlayed on a global map, as shown in \figref{overall_vs_ions}. \figref[a]{overall_vs_ions} displays the total flux of the particles detected by the apparatus. To enhance the visibility of flux variations outside the South Atlantic Anomaly (SAA), the flux values have been capped at a predefined maximum threshold. In \figref[b]{overall_vs_ions}, the map illustrates the recorded distribution of HZE ions, which are particles capable of inducing CSEEs. The data confirm that the majority of cosmic particles in low Earth orbit (LEO) cannot trigger CSEEs. Of particular interest is the absence of elevated HZE ion flux within the SAA region, which can be explained by the Van Allen inner radiation belt predominantly trapping protons and electrons, while HZE ions are not confined in the same manner.

The COTS-Capsule experiment was not designed to protect its own electronics from radiation. The spaceborne experiment operated successfully under space conditions for nearly two months, comfortably exceeding the mission’s required duration. At the end of this period, the payload ceased to function. Post-mission examination revealed that a radiation-induced CSEE caused a malfunction in the Microchip LAN9514 communication component, which is part of the RPi computer board. This failure occurred despite similar RPi boards previously operating successfully on-orbit~\cite{guertin2022raspberry, HONESS201743}. The CSEE that concluded the data-gathering mission highlights the critical importance of the COTS-Capsule apparatus in mitigating such risks for space systems that depend on COTS electronics.

\subsubsection{COTS-Capsule Emulator Logic and Algorithm}

\begin{figure}[H]%
\centering
\includegraphics[width=0.9\textwidth]{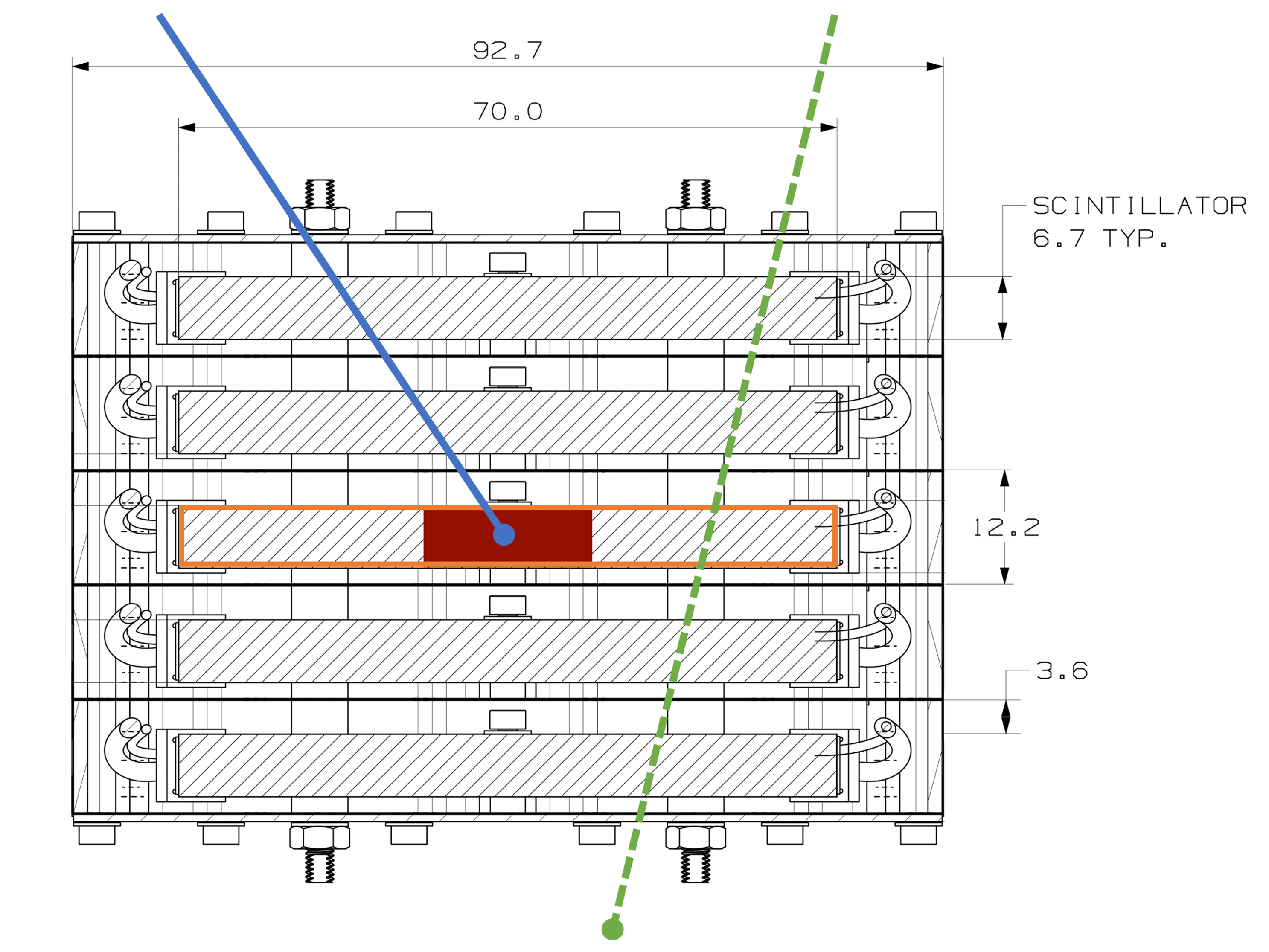}
\caption{Side view of the hodoscope, depicting the particle detector array~\cite{SSPD} within the spaceborne COTS-Capsule experiment~\cite{COTS-CAPSULE-SYSTEM}. The red rectangle, shown in the middle of the orange-framed scintillator, emulates a sensitive electronic-component. The solid blue and dashed green tracks represent particles traversing the middle particle detector, possibly triggering a CSEE. All units are in millimeters.}\label{fig:particle_telescope}
\end{figure}

The five-layered hodoscope, shown in \figref{particle_telescope}, represents the COTS-Capsule’s particle detector array and emulated CSEE-sensitive electronics. The third (middle) layer of the hodoscope emulates a radiation-sensitive electronic board containing a sensitive component, marked in red, that requires CSEE mitigation. The outermost four detectors—two on the top and two on the bottom—constitute the COTS-Capsule particle detector array. The figure illustrates two potential particle trajectories, indicated by blue and green tracks. The solid blue-track particle traverses the sensitive component, potentially triggering a CSEE, whereas the dashed green-track particle bypasses the sensitive component, thus not triggering a CSEE.

The middle detector in the array measured the energy deposited by passing particles within the emulated CSEE-sensitive electronics, providing data essential for establishing the ground truth on whether each event might have resulted in a CSEE.

\begin{figure}[H]
\centering
\includesvg[width=1\textwidth]{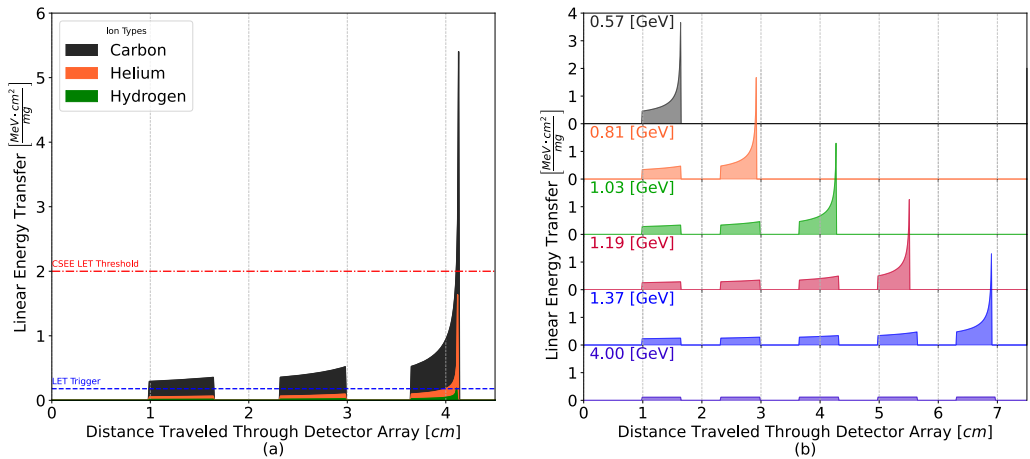}\label{fig:figure1}
\caption{SRIM simulations of the along-track linear energy transfer (LET) of particles traversing the hodoscope and detected by the particle detectors. (a) LET profiles for hydrogen, helium, and carbon ions with energies sufficient to stop in the middle scintillator. (b) LET profiles for carbon ions traversing the hodoscope with initial energies that bring the particles to stop within each of the five detectors or without stopping at all.}
\label{fig:overall1}
\end{figure}

Primary ion cosmic-particles can be categorized into three main groups: 
\begin{itemize}
    \item Light ions with an atomic number $Z \leq 2$, which are unlikely to initiate CSEEs. 
    \item HZE ions with an atomic number $Z \geq 3$, such as carbon, can induce CSEEs if they come to a complete stop within the sensitive volume of an electronic component, as shown in \figref{overall1}. 
    \item Very high atomic number ions, such as iron ions $Z \geq 26$, which can induce CSEEs even when traversing a sensitive component without coming to a complete stop.
\end{itemize}

To model and analyze particle interactions with the COTS-Capsule, we conducted SRIM simulations, as shown in \figref{overall1}. These simulations illustrate the along-track linear energy transfer (LET) of particles as they pass through the hodoscope and are detected by the particle detectors. For particles that traverse the emulated CSEE-sensitive component located within the middle detector, a decision to initiate a power cycle (CSEE mitigation) can be made based on energy measurements obtained from the surrounding four detectors.

\figref[a]{overall1} presents the along-track LET of hydrogen, helium, and carbon ions in Polyvinyl Toluene (PVT), with initial energies that bring the particles to a complete stop within the middle particle detector. 

\figref[b]{overall1} shows carbon ions traversing the hodoscope, with initial energies that either stop the particles within one of the five detectors or allow them to pass through. This simulation demonstrates that energy deposition is significantly higher when a particle stops within a detector, in particular see the LET in the third detector, whereas the energy deposition is notably lower in the scintillators through which the particle passes. By analyzing the energy absorbed by the four outermost detectors, it becomes possible to infer whether the particle may induce a CSEE in the emulated sensitive component.

The LET of a particle traversing the sensitive component, emulated by the middle detector, is estimated using data from the four outermost detectors. Specifically, for a particle detected on both sides of the emulated component, the maximum LET recorded in these outer detectors provides a conservative (worst-case) estimate of the LET within the middle detector. If this estimated LET exceeds the {\it CSEE LET Threshold}, the particle is classified as a heavy ion, such as an Iron ion, which can potentially induce a CSEE even without stopping in the sensitive component.

In cases where a particle passes through the first two detectors but does not reach the fourth detector, it is likely to have stopped within either the second particle detector or the middle detector. In this scenario, it is necessary to differentiate a particle capable of causing a CSEE (second group particle—HZE ion) from one that is not (first group particle—light ion). This differentiation is achieved by analyzing the energy deposited in the first two SSPDs and determining whether the particle LET in these detectors exceeds the predefined {\it LET Trigger} threshold. This threshold can be adjusted based on the CSEE LET Threshold specific to the sensitive component. If the LET detected by the first two SSPDs is lower than {\it LET Trigger}, the particle is classified as benign. Conversely, if it exceeds the {\it LET Trigger}, the particle is considered capable of inducing a CSEE.

For example, as shown in \figref[a]{overall1}, hydrogen and helium ions cannot initiate a CSEE in the emulated sensitive component, as their maximum LET is below the {\it CSEE LET Threshold}. In contrast, carbon ions can potentially trigger a CSEE, but only if they come to a complete stop within the emulated sensitive component. We can define $LET~Trigger=0.18~\textrm{MeV} \cdot \textrm{cm}^2 \cdot \textrm{mg}^{-1}$. By comparing the LET measured by the outer four detectors with this {\it LET Trigger} threshold, the apparatus can distinguish hydrogen and helium ions, which will not initiate a CSEE, from carbon ions and heavier ions that might.

A method to classify and differentiate potential CSEE particle events from non-CSEE events is described in Algorithm \ref{algo_decision_csee}, and works as follows. For each scintillator with available measurements, the particle’s incident location is estimated using the algorithm described in~\cite{SSPD}. Linear regression is then applied to determine the particle’s track along the hodoscope.

Based on the particle’s track and the SSPD's measurements, the LET in each scintillator can be estimated using Equation (2) in~\cite{SSPD}. This provides sufficient data to assess whether the particle passed through the sensitive component and if the estimated energy deposited within the sensitive component was high enough to warrant a power cycle.

\begin{algorithm}[H]
\caption{Particle classification and CSEE mitigation algorithm}\label{algo_decision_csee}
\begin{algorithmic}[0]
\small
\raggedright       
\algrenewcommand\algorithmicindent{1.0em} 
    \For{\textit{$SSPD\_id \in \{1,2,4,5\}$}}     \Comment{Estimate particle's location in each SSPD}

        \State $Meas\_Locs \gets Measured\_Location(\text{SSPD}[SSPD\_id])$
    \EndFor
    \State
    \State $ParticleTrack \gets \text{LinearRegression} (Meas\_Locs)$     \Comment{Use linear regression to estimate particle's track}

    \State
    \For{\textit{$SSPD\_id \in \{1,2,4,5\}$}} \Comment{Using particle's track estimate LET}
      \State $Meas\_LET[SSPD\_id] \gets \text{Meas\_LET}(\text{SSPD}[SSPD\_id], ParticleTrack)$ 
    \EndFor
    \State

\State $ChipIncidentLoc \gets ParticleTrack(\text{ChipLocation})$ \Comment{Using particle's track}\hfill\\
\hfill{to check if it passed through the chip}
    
    \State $ChipLET \gets \text{Average}(Meas\_LET)$ \Comment{Estimate particle's LET}

    \State
    \If {$ChipIncidentLoc = \text{TRUE} \And ChipLET > LET_{TRIGGER}$}
        \State $PowerCycle \gets \text{TRUE}$
    \Else {\State $PowerCycle \gets \text{FALSE}$}
    \EndIf

\end{algorithmic}
\end{algorithm}

%%%%%%%%%%%%%%%%%%%%%%%%%%%%%%%%%%%%%%%%%%%%%%%%%%%

The spaceborne experiment was subject to constraints compared to a fully implemented COTS-Capsule apparatus. To address these limitations, we applied a series of mathematical adjustments to simulate the performance of a fully operational COTS-Capsule. These adjustments account for the restricted capabilities of the current setup and enable us to extract meaningful insights into the expected behavior of the system based on the collected spaceborne data.

The first adjustment addressed the partial coverage of the emulated sensitive electronics by the hodoscope, which covers only two faces of the sensitive volume. In contrast, a fully operational COTS-Capsule apparatus would employ particle detectors on all six faces. To compensate for this, we normalized the event rate of the hodoscope to represent the geometric equivalent of a fully encapsulated configuration. Specifically, we determined the fraction of particles that traversed both the top and middle detectors in relation to the total number of events passing through the top detector. This fraction was estimated using the integral:

\begin{equation}
\frac{Flux_{\text{top \& middle}}}{Flux_{\text{top}}} = \frac{1}{\pi \cdot \text{Area}(S)}\int \int_S \Omega(x,y) \, dx \, dy \approx 44 \%, \label{eq:flux_equation}
\end{equation}
where $S$ designates the surface of the third (middle) scintillator, and $\Omega(x,y)$ represents the solid angle of the cone defined by the convex hull from the point $(x,y)$ on the surface of the third scintillator to the surface of the first (top) scintillator of the hodoscope.

This fraction was further confirmed using the formula presented in \cite[Eq.~11]{sullivan1971geometric}. Data from the COTS-Capsule experiment supported this estimate as well: the ratio of measured particle flux that traversed both the top and middle detectors to the flux that traversed only the top detector was found to be $47\%$.

The second mathematical adjustment addressed the limited dynamic range of the readout electronics used in the COTS-Capsule spaceborne experiment. This limitation restricted the detectors to measuring signal intensities within a specific range, beyond which the signals either saturated or fell below the sensitivity threshold. The system was designed to detect all ionizing particles, including light ions, while maintaining good position accuracy and LET estimation. However, this design constraint hindered the accurate assessment of the impinging position and LET for HZE ions that exceeded the dynamic range of the system ($LET\gtrsim 0.025~\textrm{MeV} \cdot \textrm{cm}^2 \cdot \textrm{mg}^{-1}$).

To unambiguously test the performance of the empirical COTS-Capsule emulator with its limited dynamic range, we were compelled to emulate a CSEE-sensitive component with a very low CSEE LET threshold. This threshold makes the component highly susceptible to CSEE, including from helium ion events, resulting in a relatively high power-cycle rate specific to the test setup. In practical scenarios, the power-cycle rate would be significantly lower.

%%=============================================================%%
%% COTS-Capsule Modelling and Simulations %%
%%=============================================================%%

\subsection{COTS-Capsule Modeling and Simulations}

We modeled the expected on-orbit performance of the COTS-Capsule apparatus aboard the ISS. Particle flux was simulated using CREME96, while particle interactions with the hodoscope were modeled in GEANT4. The results were combined and the simulated performance was analyzed using Algorithm \ref{algo_decision_csee}.

%%%%%%%%%%%%%%%%%%%%%%%%%%%%%
%
\subsubsection{Modelling the Space Environment}
%
%%%%%%%%%%%%%%%%%%%%%%%%%%%%%
%

The in-orbit radiation environment for the COTS-Capsule was modeled using the CREME96 tool, following the methodology outlined in~\cite{miller2018investigation}. The simulation included contributions from cosmic particles, solar energetic particles, and particles trapped within the Van Allen radiation belts. The orbit of the ISS was modeled for a one-year duration in 2022.

This approach generated the differential particle flux that impacted the COTS-Capsule experiment, accounting for variations in particle distributions along the ISS’s trajectory over the specified period. The goal was to estimate the fluence of various particle types as a function of the particle energy along the ISS path. Using this data, we assessed the flux of both benign and potentially harmful particles on the simulated COTS-Capsule emulator.

Using the CREME96 simulation data, we calculated the total particle flux along the ISS's orbit to be $1.19~events  \cdot sr^{-1} \cdot cm^{-2} \cdot s^{-1}$, which represents the average value shown in \figref{overall_vs_ions}a and accounts for the increased flux over the SAA. The simulations also indicate an HZE ion flux of $5 \cdot 10^{-4}~events  \cdot sr^{-1} \cdot cm^{-2} \cdot s^{-1}$, aligning with \figref{overall_vs_ions}b, where HZE ions are defined as those with an estimated LET of at least $0.02~MeV \cdot cm^2 \cdot mg^{-1}$.

%%%%%%%%%%%%%%%%%%%%%%%%%%%%%%
%
\subsubsection{Modelling Particle Interactions with the Hodoscope}
%
%%%%%%%%%%%%%%%%%%%%%%%%%%%%%%%
%

The COTS-Capsule hodoscope was modeled using the GEANT4 simulation toolkit. The model includes the hodoscope structure, scintillators, and SiPM sensors.

A total of 40,000 particle events were simulated, including common benign particles such as hydrogen and helium ions, as well as potentially harmful HZE ions prevalent in the space environment. The simulations involved directing particles at the hodoscope with varying energies and trajectories. As shown in \figref{particle_telescope} and \figref{overall1}, CSEEs are most likely when the HZE ions come to a complete stop within the sensitive component. To ensure sufficient representation of these rare but critical events, we simulated a broad range of CSEE-inducing HZE ion events. The resulting data were then normalized to match the on-orbit particle flux and trajectory distributions.

The data from each event include (1) particle characteristics, trajectory, and energy; (2) interaction details within the scintillators, including the energy deposited per scintillator and the number of photons generated; and (3) the number of photons reaching each SiPM sensor.

By analyzing the impact position and maximum LET of each particle within the emulated sensitive component, as determined by GEANT4, we established the ground truth for potential CSEEs. The performance of the simulated COTS-Capsule emulator was then evaluated by processing the photon counts recorded by SiPMs on the four outer SSPDs (excluding the middle detector) using Algorithm \ref{algo_decision_csee}.

%%%%%%%%%%%%%%%%%%%%%%%%%%%%%%%%
%
\section{COTS-Capsule Emulator: Results and Performance}
%
%%%%%%%%%%%%%%%%%%%%%%%%%%%%%%%%
%

The performance of the COTS-Capsule is defined by two key parameters: the acceptable power-cycle rate and the tolerable level of missed CSEEs. By mitigating CSEEs, the COTS-Capsule extends the average lifetime of sensitive electronics in space. The False-Negative (FN) rate, which represents missed CSEEs, determines the increase in average lifetime, while the Predicted Positives (PP) dictate the power-cycle rate.

We consider a true positive rate of $95\%$ acceptable, which allows $5\%$ of CSEE events to go undetected. This true positive rate implies that the COTS-Capsule emulator can, on average, extend the operational lifetime of sensitive electronics in space by a factor of twenty.

A tolerable power-cycle rate depends on the criticality of the electronics’ function and the requirements of the mission timing. For example, in image processing tasks, a Mars rover using a high-performance GPU for navigation can tolerate periodic power cycles, as its mild real-time constraints allow time for system recovery. 
In contrast, an imaging satellite performing on-orbit classification may demand stricter real-time performance to deliver results while passing over a target area. In this case, redundant configurations, such as parallel GPUs with hot redundancy, can ensure continuous operation with minimal disruption.

The mitigation scheme can be fine-tuned based on the CSEE sensitivity of the electronic components. The scheme can be adjusted accordingly if precise testing has established exact thresholds for specific components. Conversely, if CSEE sensitivity data is unavailable, a conservative threshold of $LET_{th} \gtrsim 1.2~\textrm{MeV} \cdot \textrm{cm}^2 \cdot \textrm{mg}^{-1}$ can be applied, as this minimum threshold is broadly applicable to HZE ion events. Utilizing calibrated thresholds reduces the frequency of power cycling, resulting in a more efficient mitigation approach.

%%%%%%%%%%%%%%%%%%%%%%%%%%%%%%%%%%%%%%%
%
\subsection{COTS-Capsule Emulator: Simulated Performance}
%
%%%%%%%%%%%%%%%%%%%%%%%%%%%%%%%%%%%%%%%
%

\begin{figure}[H]
%\centering
\begin{flushleft}
\includegraphics[width=1\textwidth]{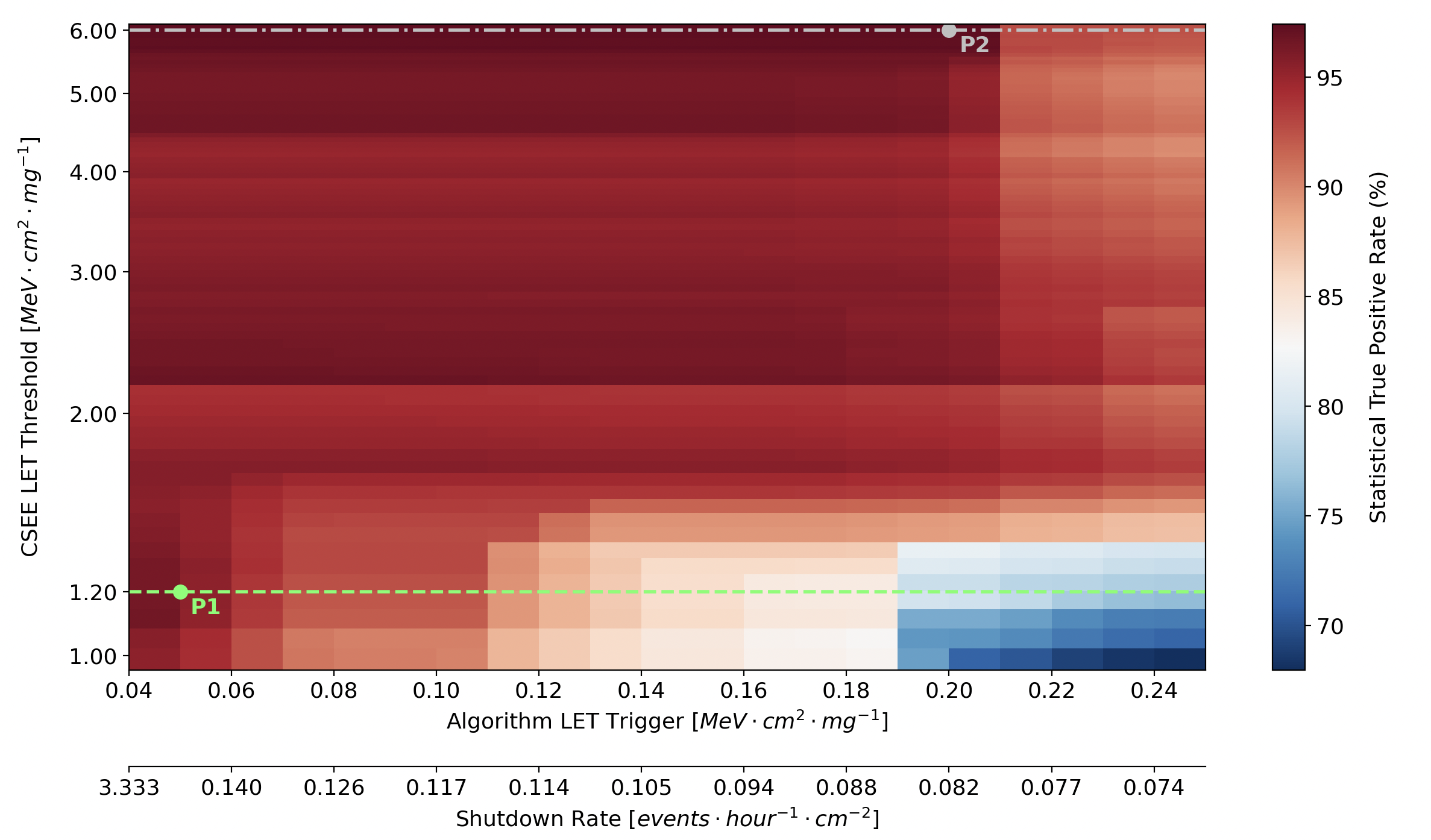}
\end{flushleft}
\caption{A heat map illustrating the simulated performance of the COTS-Capsule apparatus in orbit. The y-axis shows the susceptibility of the electronics to CSEE, indicated as the {\it CSEE LET Threshold}. The x-axis represents the {\it LET trigger} parameter applied in Algorithm \ref{algo_decision_csee}, with the corresponding power-cycle rate shown on the secondary x-axis. Heat map colors convey the apparatus’s statistical true positive rate. Two possible approaches are suggested: {\it Conservative susceptibility} - Assumes the electronics are susceptible to all HZE ions. This approach is suitable if CSEE susceptibility is unknown or untested with $\text{CSEE LET threshold} \gtrsim 1.2~MeV \cdot cm^2 \cdot mg^{-1}$, marked by a green dashed line; {\it Simplified risk assessment} - Assumes the electronics have passed the simplified risk assessment for CSEE, as per \cite{o1997risk}, with $\text{CSEE LET threshold} \gtrsim 6~MeV \cdot cm^2 \cdot mg^{-1}$, shown by a gray dot-dashed line. A 95~$\%$ statistical true positive rate is marked as P1 and P2 on the lines representing these two threshold approaches.}
\label{fig:Simulations_Power_Cycle_Rate}
\end{figure}

\figref{Simulations_Power_Cycle_Rate} presents a heat map of the simulated on-orbit performance of the COTS-Capsule emulator aboard the ISS. The y-axis represents the CSEE LET threshold, indicating the susceptibility of the electronics to CSEE, while the x-axis shows the LET trigger parameter used in Algorithm \ref{algo_decision_csee}. The secondary x-axis provides the corresponding power cycle rate. The colors of the heat map indicate the statistical true positive rate of the apparatus, defined as the percentage of correctly identified potential CSEE events out of all potential CSEE events.

Two mitigation approaches are suggested based on different levels of CSEE susceptibility. The first approach, {\it Conservative susceptibility} assumes the electronics are susceptible to all HZE ions and is suitable for untested or unknown susceptibility, with $\text{CSEE LET threshold} \gtrsim 1.2~\textrm{MeV} \cdot \textrm{cm}^2 \cdot \textrm{mg}^{-1}$. The second approach, {\it Simplified risk assessment} applies to components that were tested and passed the simplified CSEE risk assessment test, as outlined in \cite{o1997risk}, allowing for a less stringent $\text{CSEE LET threshold} \gtrsim 6~\textrm{MeV} \cdot \textrm{cm}^2 \cdot \textrm{mg}^{-1}$.

Both approaches include operating points, marked as P1 and P2, corresponding to a statistical true positive rate of $95\%$. For untested electronics, the conservative $\text{CSEE LET threshold} \gtrsim 1.2~\textrm{MeV} \cdot \textrm{cm}^2 \cdot \textrm{mg}^{-1}$, with a true positive rate of $95\%$, requires setting $LET_{trigger}=0.05~\textrm{MeV} \cdot \textrm{cm}^2 \cdot \textrm{mg}^{-1}$, leading to a power-cycle rate of once every seven hours per cm$^2$ of sensitive component area (P1). For components that were tested and passed the simplified CSEE risk assessment test, the less stringent $\text{CSEE LET threshold} \gtrsim 6~\textrm{MeV} \cdot \textrm{cm}^2 \cdot \textrm{mg}^{-1}$, with a true positive rate of $95\%$, requires setting $LET_{trigger}=0.2~\textrm{MeV} \cdot \textrm{cm}^2 \cdot \textrm{mg}^{-1}$ resulting in a power-cycle rate of once every twelve hours per cm$^2$ of sensitive component area (P2).

%%%%%%%%%%%%%%%%%%%%%%%%%%%%%%%%
%
\subsection{COTS-Capsule Emulator: Empirical Performance in Orbit} 
%
%%%%%%%%%%%%%%%%%%%%%%%%%%%%%%%%
%
The COTS-Capsule experiment demonstrated the ability to reconstruct light ion trajectories. However, the limited dynamic range of the readout electronics restricts trajectory reconstruction for HZE ions when two or more SiPMs become saturated by the large number of photons generated as these ions traverse the scintillator.

To establish the ground truth for potential CSEEs, we analyzed data from the entire hodoscope, including the middle detector. Particles capable of inducing a CSEE deposit enough energy to saturate all the detectors they traverse, preventing the determination of their LET and precise impact position. Nevertheless, these particles can be identified as CSEE-capable based on the detectors they saturate and those they do not. A particle is classified as CSEE-capable if it stops in the middle detector, as shown by the blue particle in \figref{particle_telescope}. In this case, the top two detectors register the particle, while the bottom two do not, or vice versa. Using this method we obtain the number of CSEE particle flux per hour. Then, using Equation (\ref{eq:flux_equation}) we normalize the flux per area to receive the ground truth of $2.3~events \cdot cm^{-2} \cdot hour^{-1}$.

Higher atomic-number HZE ions, such as iron, can induce CSEEs without fully stopping in the sensitive electronics. Including these ions increases the total potential CSEE flux by $30\%$, resulting in a total flux of $3~events \cdot cm^{-2} \cdot hour^{-1}$.

The COTS-Capsule emulator uses data from the four outer detectors to assess the power-cycle rate with Algorithm \ref{algo_decision_csee}. To address the limited dynamic range of the spaceborne experiment, a modified version, Algorithm \ref{algo_decision_csee_sat}, was developed.

Algorithm \ref{algo_decision_csee_sat} identifies a particle as CSEE-capable if it generates enough photons to saturate the detectors it traverses and stops in the middle detector. While data from the middle detector are excluded, the behavior of the particle is inferred from the outer detectors. Specifically, a particle is classified as CSEE-capable if it is detected by the top two detectors but not the bottom two, or vice versa.

\begin{algorithm}[H]
\caption{Particle Classification and CSEE Mitigation Algorithm}
\label{algo_decision_csee_sat}
\begin{algorithmic}[0]
\small
\raggedright          
\algrenewcommand\algorithmicindent{1.0em} 

    \For{\textit{$SSPD\_id \in \{1,2,4,5\}$}} 
        \State $Meas\_LET[SSPD\_id] \gets \text{FALSE}$
        \For{\textit{$idx \in \{1,2,3,4\}$}}
            \State $sat\_count \gets 0$
            \State $measured \gets 0$
            \If {$detector[idx] \geq threshold_{saturation}$}
                \State $sat\_count \gets sat\_count + 1$ \Comment{Count saturated SiPMs}
            \EndIf
            \If {$detector[idx] \geq 0$}
                \State $measured \gets measured + 1$
            \EndIf
        \EndFor
       \If {$measured \geq 2$}
          \State $Meas\_LET[SSPD\_id] \gets \text{TRUE}$
          \Comment{Detector is saturated}
       \EndIf

        \If {$sat\_count = 4$}
         \State $Meas\_LET[SSPD\_id] \gets \text{HIGH}$
        \EndIf
    \EndFor

    \If {$Meas\_LET[1] = \text{HIGH} \textbf{ and } Meas\_LET[2] = \text{HIGH} \textbf{ and } Meas\_LET[4] = \text{FALSE}$}
        \State $PowerCycle \gets \text{TRUE}$ \Comment{Potential CSEE - powercycle}
    \ElsIf {$Measured\_LET[4] = \text{HIGH} \textbf{ and } Meas\_LET[5] = \text{HIGH} \textbf{ and } Meas\_LET[2] = \text{FALSE}$}
        \State $PowerCycle \gets \text{TRUE}$ \Comment{Potential CSEE - powercycle}
    \Else
        \State $PowerCycle \gets \text{FALSE}$
    \EndIf

\end{algorithmic}
\end{algorithm}

Applying Algorithm \ref{algo_decision_csee_sat} to the on-orbit empirical data, the power-cycle rate is $4.2~events \cdot cm^{-2} \cdot hour^{-1}$. Including very high atomic number ions increases the rate to $5.5~events \cdot cm^{-2} \cdot hour^{-1}$.

%%%%%%%%%%%%%%%%%%%%%%%%%%%%%%%%
\subsection{Comparison of Simulated and Empirical Results}
%%%%%%%%%%%%%%%%%%%%%%%%%%%%%%%%%

The simulated COTS-Capsule emulator generates a comprehensive dataset across various operating points, while the empirical emulator validates these results with real-world data. A comparison of the two shows that the theoretical model accurately represents practical performance at the operating point defined by the empirical emulator.

To simulate the limited dynamic range, we model events in the simulation and calculate the number of photons reaching the SiPMs. We then compare the actual signal, including its saturation, to the photon counts measured by the simulated SiPMs. This process is used to calibrate the two models against each other.

In the experimental setup, the maximum detected LET for a particle passing through the center of the detector was $0.025~MeV \cdot cm^2 \cdot mg^{-1}$. Simulations indicated that this LET corresponds to $3500-4500$ photons reaching each SiPM, resulting in saturation of the readout electronics. These saturation thresholds were applied to the simulated data to reflect the results that would be measured in the empirical COTS-Capsule setup.

Algorithm \ref{algo_decision_csee_sat} was applied to the saturation-adjusted dataset from the simulated COTS-Capsule emulator. The ground truth, representing the optimal power-cycle rate derived from the simulated emulator, was $2.8-3.8~events \cdot cm^{-2} \cdot hour^{-1}$, consistent with the empirically determined ground truth value of $3~events \cdot cm^{-2} \cdot hour^{-1}$. Furthermore, the average power-cycle rate derived from the simulation, $4.8-6.7~events \cdot cm^{-2} \cdot hour^{-1}$, aligns with the power-cycle rate of $5.5~events \cdot cm^{-2} \cdot hour^{-1}$ obtained from the empirical COTS-Capsule emulator using Algorithm \ref{algo_decision_csee_sat}.

Notably, the limited dynamic range of the readout electronics in the empirical COTS-Capsule emulator necessitates testing at an operating point that emulates electronic components sensitive to light ions. This operating point leads to a higher power-cycle rate than is required for the apparatus. Enhancing or fine-tuning the dynamic range of the readout electronics will enable operation at the P1 and P2 points discussed in \figref{Simulations_Power_Cycle_Rate}, significantly reducing the power-cycle rate.

%%==============================================%%
%%      Conclusions                 %%
%%==============================================%%

\section{Conclusions}

The COTS-Capsule is a non-intrusive method for mitigating CSEEs in spaceborne electronics.
This paper details the concept, underlying physics, and engineering design, supported by results from a dedicated spaceborne experiment conducted aboard the ISS. The experiment emulated the design and behavior of the COTS-Capsule apparatus, with performance analyzed and compared to simulations.

The COTS-Capsule safeguards sensitive electronics by enclosing them within an array of particle detectors that monitor cosmic particle interactions. A power-cycling algorithm is activated only when particles capable of causing CSEEs are detected, preventing damage while minimizing operational disruptions.

Simulations using CREME96 and GEANT4, along with empirical data from the ISS experiment, confirmed the system’s performance, showing strong alignment between the results.

The results demonstrate that the COTS-Capsule effectively mitigates CSEEs, achieving a power-cycle rate of no more than once every seven hours per square centimeter of sensitive area. At this suggested operational point, the system extends the lifetime of electronics in space by a factor of 20 while ensuring reliable performance. This power-cycle rate underscores the suitability of the system for a wide range of space systems and missions.

By enabling the use of high-end commercial off-the-shelf (COTS) electronics in space without requiring custom designs or intrusive modifications, the COTS-Capsule offers a cost-effective and time-efficient solution. These results confirm its feasibility and broad applicability across diverse space missions.

\section*{Acknowledgements}\label{sec11}

We appreciate the assistance of the staff at Tel Aviv University. We thank Israel Aerospace Industries for their support in designing the COTS-Capsule experiment. We are grateful for the Ramon Foundation for managing the Rakia mission.
Acknowledgment to ISS-National Lab/NASA via the implementation partner Nanoracks LLC supported by Nanoracks Space Outpost Europe srl.
Special thanks go to the staff at Soreq Nuclear Center for reviewing and commenting on the initial design and Dr. Sari Katz for providing CREME96 simulation results for the ISS environment. We thank CAEN and Bern University for their technical support. We thank Yuri Orlov and Dr. Dolev Bashi for their work on the COTS-Capsule mission to the ISS. We appreciate Aviv Erlich's work on creating a large GEANT4 simulation dataset and taking the COTS-Capsule algorithm one step forward. We thank Aron Rubin for reproducing the Teensy\textsuperscript \textregistered \ LC 3D model.

\bibliographystyle{elsarticle-num}
\bibliography{sn-bibliography}    

\newpage

\end{document}